\documentclass{aa}

\newcommand{\kms}{{\,km\,s$^{-1}$}}
\newcommand{\sci}{{\,{\sc i}}}

\newcommand{\vsini}{{$v\sin i$}}

\begin{document}

   \thesaurus{06         %
              (08.01.2;  
               08.09.1;  
               08.09.2;  
               08.12.1;  
               08.19.6:)}
   \title{Study of FK~Comae Berenices}
   \subtitle{II. Spot evolution from 1994 to 1997\thanks{Based on the observations obtained at the Nordic Optical Telescope, Observatorio Roque de los Muchachos, La Palma, Canary Islands, Spain; the National Astronomical Observatory, Rozhen, Bulgaria; the Kitt Peak National Observatory, USA; the Automatic Photometric Telescope, Phoenix 10, Arizona, USA.}}

   \author{H. Korhonen \inst{1}
   \and S.V. Berdyugina \inst{1}
   \and T. Hackman \inst{2}
   \and K.G. Strassmeier\inst{3}
   \and I. Tuominen \inst{1}}

   \offprints{H.\ Korhonen: heidi.korhonen@oulu.fi}

\institute{ Astronomy Division, P.O. Box 3000, 
            FIN-90014 University of Oulu, Finland
   \and Observatory, P.O. Box 14,  FIN-00014 University of Helsinki, Finland
   \and Institute for Astronomy, University of Vienna, T{\"u}rkenschanzstr.\ 17, A-1180 Vienna, Austria}

   \date{Received; accepted }

   \maketitle

\begin{abstract}

We present new surface (Doppler) images of the late-type single giant FK~Com for June-July 1996, July-August 1996, April 1997 and June 1997. These images are compared with the previously published images from 1994 and 1995. The consecutive maps are cross-correlated to see the possible migration of the spots and the effects of differential rotation. The cross-correlation confirms an average longitudinal spot migration of $0.22\pm0.03$ in phase within a year. This movement is probably an artifact caused by a difference between the accepted rotation period and the real photometric period for these years. If this is true, then the photometric rotation period for these years is $2\fd4037\pm 0.0005$. Measurements from these four years and six maps limit the surface differential rotation to $\alpha=0.0001\pm0.0002$, where $\alpha$ is the difference between polar and equatorial angular velocities relative to the equatorial angular velocity.

\end{abstract}

\begin{table*}
\caption{Spectroscopic observations of FK~Com obtained between 1996 and 1997. Phases were calculated with the ephemeris given in Eq.~\ref{eq:hjd}.}
\begin{center}
\begin{tabular}{lllllllll}\hline
HJD &Phase &S/N &HJD &Phase &S/N &HJD &Phase & S/N  \\
2450000+&    &   &2450000+&   &  &2450000+ & &  \\ \hline 
\multicolumn{3}{c}{\it June-July 1996, Rozhen} & \multicolumn{3}{c}{\it  April 1997, KPNO} & \multicolumn{3}{c}{\it June 1997, NOT} \\
264.32 & 0.62 & 173 &545.81 &0.90 &295 &617.40 &0.72 & 288 \\ 
265.37 & 0.06 & 264 &546.80 &0.31 &360 &617.43 &0.74 & 217 \\
266.31 & 0.45 & 228 &547.80 &0.73 &275 &619.38 &0.55 & 202 \\ 
267.32 & 0.87 & 338 &548.82 &0.15 &284 &619.41 &0.56 & 308 \\
268.31 & 0.28 & 275 &549.85 &0.58 &304 &620.40 &0.97 & 337 \\
\multicolumn{3}{c}{\it July-August 1996, NOT} &550.76 &0.96 &309 &621.39 &0.39 & 194 \\
293.39 & 0.73 & 261 &551.79 &0.39 &327 &621.43 &0.40 & 254 \\ 
294.38 & 0.14 & 300 &552.77&0.81 &351 &622.41 &0.81 & 324 \\
295.39 & 0.57 & 214 &553.79 &0.22 &381 &623.41 &0.23 & 344 \\
297.38 & 0.40 & 238 &554.80 &0.64 &334 &624.39 &0.64 & 218 \\
298.39 & 0.82 & 331 & & & &624.42 &0.65 & 289 \\
299.38 & 0.23 & 265 & & & &625.40 &0.06 & 340 \\
 & & &       &      &     &626.39 &0.47 & 185 \\
 & & &       &      &     &626.42 &0.48 & 259 \\
 & & &       &      &     &627.39 &0.89 & 199 \\
 & & &       &      &     &627.42 & 0.90 & 247 \\ \hline
\end{tabular}
\end{center}
\label{spect}
\end{table*}

\section{Introduction}

FK~Com (HD117555) is the prototype of the small group of rapidly rotating single G-K giants, the FK~Com-type stars (Bopp \& Rucinski \cite{bopp2}; Bopp \& Stencel \cite{bopp3}; Bopp \cite{bopp1}). FK~Com itself is classified as a single G5~II-G8~III giant with a \vsini\ of $162.5\pm3.5$\kms (Huenemoerder et al.\ \cite{hue}). Our recent study gives spectral type G5~III and \vsini=155\kms (Korhonen et al.\ \cite{kor}, hereafter referred to as Paper I).

Small variations in the visual magnitude of FK~Com of $0\fm1$ with a period of $2\fd412$ were first reported by Chugainov (\cite{chu1}). These variations were later interpreted to be caused by asymmetrically distributed spots (Bopp \& Rucinski \cite{bopp2}). Dorren et al.\ (\cite{dor}) determined the temperature of these spots to be $\sim 600-800$~K cooler than the unspotted surface. 

Jetsu et al.\ (\cite{jetsu2}) determined a photometric period of 2.4002466 $\pm0^{\rm d}.0000056$ and discovered a switch of activity between two active longitudes 180 degrees apart (``flip-flop''). A summary of all available photometry up to 1996 was presented by Strassmeier et al. (\cite{str:bar}). Four months of phase resolved and continuous photometry from 1996 was presented in Strassmeier et al. (\cite{apt}) and further data and rotation periods can be found in Strassmeier et al. (\cite{str:ser}) from the 1996/97 observing season.

FK~Com also shows spectral peculiarities, of which some were first reported by Merrill (\cite{mer}). The peculiarities include a strong and variable H$\alpha$ emission (e.g.\ Ramsey et al.\ \cite{ram}), strong chromospheric UV emission (Bopp \& Stencel \cite{bopp3}) and high X-ray luminosity (Walter \cite{wal}). Similar spectral characteristics as in the H$\alpha$-line can also be seen in the other Balmer lines, the effects are just less extreme. Based on the H$\alpha$ and H$\beta$ fluxes, Huenemoerder et al.\ (\cite{hue}) have proposed that the emission seen in the spectra of FK~Com arises from structures similar to the solar prominences. The work by Oliveira \& Foing (\cite{oliv}) supports this interpretation. 

The stellar surface structures can be studied in more detail by using surface (Doppler) imaging techniques. The main idea in surface imaging is to trace distortions appearing in the observed line profiles due to the presence of spots on the stellar disk and moving due to stellar rotation. This is an ill-posed inversion problem and there are several methods developed for solving these problems, for example the Maximum Entropy Method (Vogt et al.\ \cite{vogt2}), Tikhonov Regularization (Goncharsky et al.\ \cite{gon}; Piskunov \cite{pisk1}) and the Occamian Approach (Berdyugina \cite{ber2}). 

Tikhonov regularization has previously been used for mapping the surface temperature of FK~Com for the years 1989 (Piskunov et al.\ \cite{pisk3}) and 1994 \& 1995 (Paper I). The images for the years 1994 and 1995 show an active region with two spots with very similar substructures and surface temperatures. In both cases the photometric minimum was caused by two different groups of spots, not only one spot or one spot group. In the current paper new surface temperature maps are presented for June-July 1996, July-August 1996, April 1997 and June 1997 and compared with the previously obtained maps for the summers of 1994 and 1995. The consecutive maps are cross-correlated to obtain the possible spot movement and the effects of differential rotation. In the case of the June 1997 map the position of the surface structures is also compared with the observations of the chromospheric activity analysed by Oliveira \& Foing (\cite{oliv}).

\section{Observations}

A summary of the new spectroscopic observations used can be found in Table~\ref{spect}. The spectroscopic and photometric observations for the years 1994 and 1995 were described in Paper I.

High-resolution spectral observations of FK~Com were obtained with the Nordic Optical Telescope (NOT, La Palma) and the SOFIN \'echelle spectrograph with the medium resolution camera during the summers 1996 and 1997. Typically, the \'echelle spectra consisted of 14 orders and were centred at 6427 {\AA}. In 1996, the slit width was 123~$\mu {\rm m}$, providing a resolution (${\lambda}/{\Delta\lambda}$) of 51~000. The observations consist of 12 spectra taken during different nights between the 28th of July and the 3rd of August. The observations taken during the same night were averaged to increase the signal-to-noise ratio of the individual phases. In 1997, the slit width was 81~$\mu {\rm m}$, providing a resolution (${\lambda}/{\Delta\lambda}$) of 76~000. The observations consist of 32 spectra taken during different nights between the 17th and the 27th of June. Some of the consecutive observations were averaged. The observations were combined only when they were obtained immediately after each other and the total time spent was between 40 and 90 minutes. This does not cause any significant phase smearing.

For the year 1996 another set of observations was obtained at the National Observatory in Rozhen, Bulgaria with the coud{\'e} spectrograph at the 2m RCC-telescope. For these observations, the resolution (${\lambda}/{\Delta\lambda}$) was $\sim 30~000$. The observations consist of 10 spectra taken during different nights between the 29th of June and the 3rd of July. The spectra taken during the same night have been averaged.

All the spectroscopic observations obtained at NOT and Rozhen were reduced using the 3A reduction package (Ilyin \cite{ilyin}). The reductions included bias subtraction, flat fielding, scattered light removal and order extraction. The wavelength scale was calibrated using  Th-Ar comparison spectra obtained immediately before and after the stellar exposures. The zero point of the wavelength scale was adjusted with atmospheric lines. This was done in order to remove small shifts between stellar and comparison spectrum images.

FK~Com was also observed in April 1997 at the Kitt Peak National Observatory (KPNO) with the 0.9-m coud\'e feed telescope during 10 consecutive nights from April 4--14.  The Ford 3k-CCD detector was employed together with grating~A, camera~5, the long collimator, and a 280-$\mu$m slit to give a resolving power of 27,000 at 6500 \AA , i.e. 0.25~\AA \ as judged from the full width at half maximum (FWHM) of unblended Th-Ar comparison-lamp lines. The useful wavelength range was 300~\AA \ and the exposure time was set to 4000~sec. The analog-to-digital units of 20,000 in the continuum correspond to a signal-to-noise ratio of around 300:1 (see Table~\ref{spect}).  Forty nightly flat fields were co-added and used to remove the pixel-to-pixel variations in the stellar spectra. The Ford CCD shows no signs of fringing around 6400--6700 \AA \ and no attempts were made to correct for it other than the standard flat-field division. Continuum fitting with a very low-order polynomial was sufficient to find an excellent continuum solution. The velocity zero point was obtained by observing radial-velocity standard stars at least once per night. The KPNO reductions were done with the Image Reduction and Analysis Facility (IRAF) distributed by KPNO/NOAO.

For all the spectroscopic observations taken during the years 1996 and 1997, we also have simultaneous or nearly simultaneous photometric observations in B and V from the Automatic Photometric Telescope (APT), Phoenix 10, Arizona. For 1996, we have used 23 observations taken between the 27th of May and the 30th of June. For April 1997 we have used 39 observations taken between the 28th of March and the 27th of April, and for June 1997 30 observations obtained between the 1st of June and the 6th of July. Observations with mean errors larger than $0\fm02$ were automatically excluded from the data set.

For the phase calculation in this work we have used the ephemeris obtained from 25 years of photometric observations: 
\begin{equation}
HJD= 2,439,252.895 + (2\fd4002466\pm0\fd0000056)E, 
\label{eq:hjd}
\end{equation}
referring to a photometric minimum calculated by Jetsu et al.\ (\cite{jetsu2}, \cite{jetsu3}).

\section{Temperature mapping}

The stellar parameters adopted for surface imaging can be found in Table~\ref{para}. More details on the selection of the parameters are given in Paper I. For this work the value of the microturbulence and the effective temperature were slightly changed from the values mentioned in Paper I to better fit the line profiles and the B-V colour. 

For FK~Com, \vsini\ =155\kms\ was determined from the 1995 data using Fourier transforms, as described in Paper I. For the present work more tests for finding the best \vsini\ were done. When calculating temperature images for the June 1997 observations, the smallest deviation between the model and the observations was achieved with \vsini=159\kms (the difference between deviations obtained with these two \vsini\ values was 0.09\%). Values larger than this also gave a small deviation, but then a belt of cool spots appeared on the equator in the map, indicating a too large a value of \vsini. The tests also showed that with such rapid rotation lowering \vsini\ by 4\kms\ does not have significant effects on the temperature maps themselves. For this reason and for the consistency, it was decided to use \vsini\ = 155\kms\ also in the inversions for the years 1996--1997, it being the value originally used for the 1994 and 1995 maps. 

\begin{table}
\caption{Adopted values of the stellar parameters for surface imaging for the years 1996 and 1997.}\label{stellar}
\begin{center}
\begin{tabular}{ll}\hline
Parameter & Adopted value\\ \hline
T$_{\rm eff}$ (unspotted)	& 5000~K\\
$\log g$         		& 3.5 \\
Period         			& $2{\hbox{$.\!\!^{\rm d}$}}4002466$ \\ 
$v\sin i$        		& 155 \kms  \\
Inclination      		& $60^{\circ}$ \\
Microturbulence			& 1.0 \kms \\ 
Macroturbulence			& 2.0 \kms \\ \hline
\end{tabular}
\label{para}
\end{center}
\end{table}
\vspace{1cm}

In the inversion procedure the observed line profiles are compared to the model line profiles. Local line profiles were calculated with the code by Berdyugina (\cite{ber1}), which includes calculations of opacities in the continuum and in atomic and molecular lines. Number densities of atoms and molecules were calculated under the assumption of dissociative equilibrium. Atomic line parameters were obtained from VALD (Piskunov et al.\ \cite{pisk4}; Kupka et al.\ \cite{kupka}), while molecular line parameters were calculated as was described by Berdyugina et al.\ (\cite{ber3}). LTE stellar model atmospheres from Kurucz (\cite{kur}) were used.

The spectral line parameters for the calculations used in this work were slightly changed from the ones used in Paper I. This was due to the constant development of the VALD database. The new model fits the observed spectra without any changes in the $\log(gf)$ values, except for the Ca\sci\ 6439~{\AA} line, where the value was changed to $\log(gf)=+0.250$.

The local line profiles were calculated for 20 values of $\mu=\cos\theta$ from the disk centre to the limb. Spectra were calculated for temperatures ranging from 3500~K to 6000~K in steps of 250~K and with solar abundances.

\section{New temperature maps}

The inversions were done for all the lines in the spectral region between 6416~{\AA} and 6444~{\AA} using the Tikhonov Regularization method code, INVERS7, written by N.\ Piskunov and modified by T.\ Hackman (Hackman et al.\ \cite{hack}).

\begin{figure*}
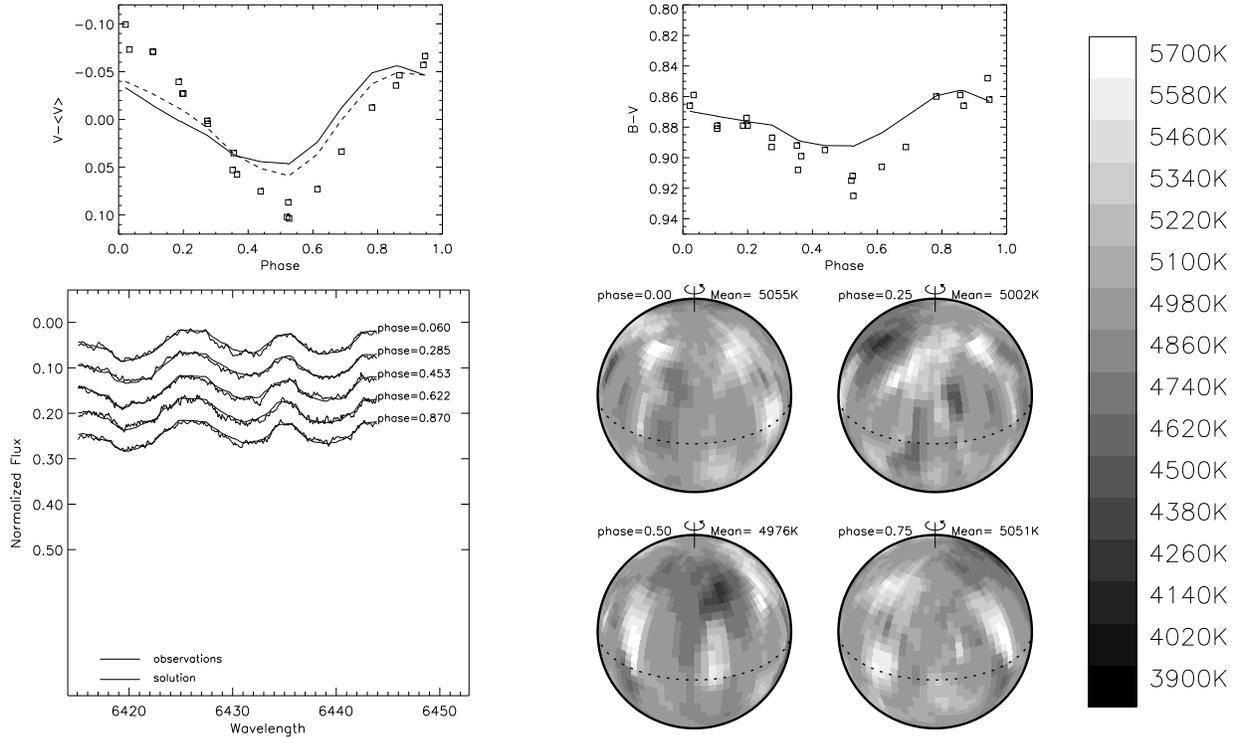

\setlength{\unitlength}{1mm}
\begin{picture}(0,98)           
\put(10,-3){\begin{picture}(0,0) \includegraphics{ms9885.f1a} \end{picture}}
\put(65,-13){\begin{picture}(0,0) \includegraphics{ms9885.f1b} \end{picture}}
\put(90,58){\begin{picture}(0,0) \includegraphics{ms9885.f1c} \end{picture}}
\put(15,58){\begin{picture}(0,0) \includegraphics{ms9885.f1d} \end{picture}}
\end{picture}
\caption{The surface temperature map of FK~Com for June-July 1996 obtained with the Tikhonov regularization from the 6416--6444 {\AA} region with \vsini=155\kms. Based on data from Rozhen. A grid of 40 latitudes and 80 longitudes across the stellar surface is used in the map. Calculated and observed spectral lines are shown by thick and thin lines, respectively. Photometric observations are plotted with squares, and curves calculated from the map are presented by lines. The dashed line in the V curve plot is the V curve calculated from the map after the cool spots at the phase interval 0.0--0.3 have been removed from the image, see the discussion in Sect.~\ref{r96}.}
\label{map96r}
\end{figure*}

\begin{figure*}
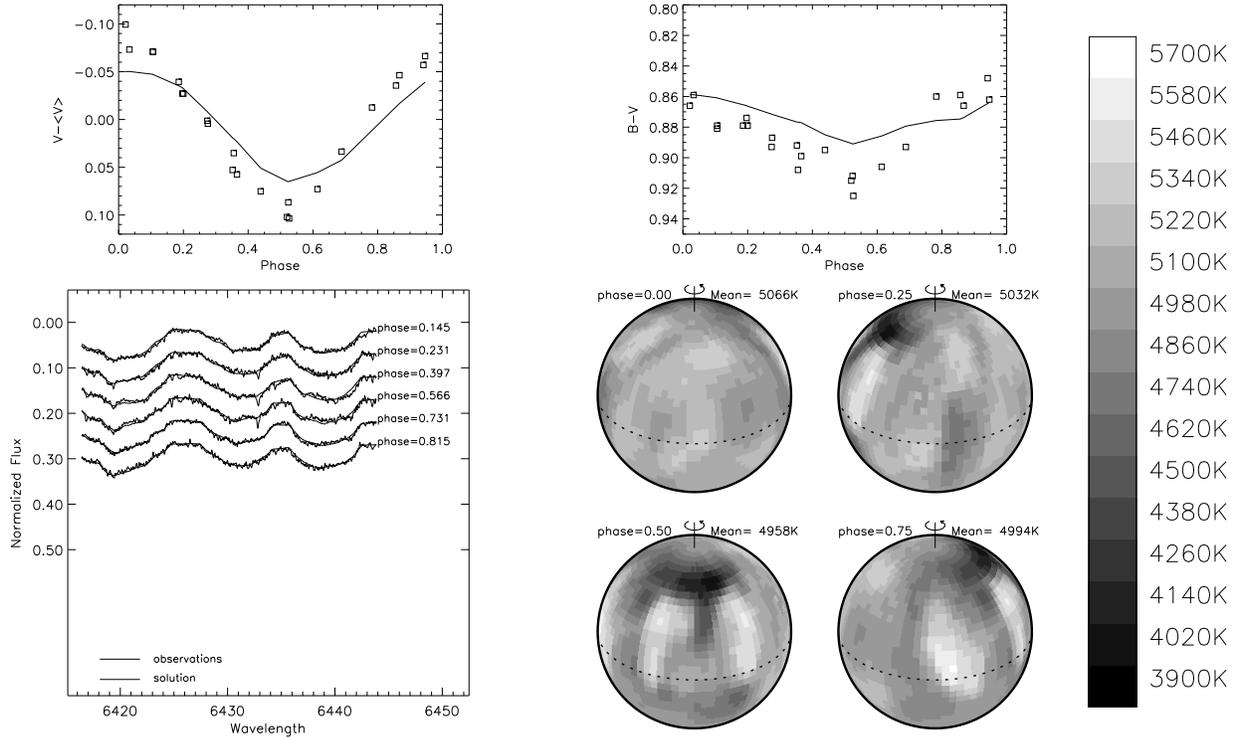

\setlength{\unitlength}{1mm}
\begin{picture}(0,98)           
\put(10,-3){\begin{picture}(0,0) \includegraphics{ms9885.f2a} \end{picture}}
\put(65,-13){\begin{picture}(0,0) \includegraphics{ms9885.f2b} \end{picture}}
\put(90,58){\begin{picture}(0,0) \includegraphics{ms9885.f2c} \end{picture}}
\put(15,58){\begin{picture}(0,0) \includegraphics{ms9885.f2d} \end{picture}}
\end{picture}
\caption{The surface temperature map of FK~Com obtained for July-August 1996. Based on data from NOT. All the other parameters are as in Fig~\ref{map96r}.}
\label{map96s}
\end{figure*}

\begin{figure*}
\setlength{\unitlength}{1mm}
\begin{picture}(0,100)           
\put(10,-3){\begin{picture}(0,0) \includegraphics{ms9885.f3a} \end{picture}}
\put(65,-13){\begin{picture}(0,0) \includegraphics{ms9885.f3b} \end{picture}}
\put(90,58){\begin{picture}(0,0) \includegraphics{ms9885.f3c} \end{picture}}
\put(15,58){\begin{picture}(0,0) \includegraphics{ms9885.f3d} \end{picture}}
\end{picture}
\caption{The surface temperature map of FK~Com obtained for April 1997. Based on data from KPNO. All the other parameters are as in Fig~\ref{map96r}.}
\label{map97a}
\end{figure*}

\begin{figure*}
\setlength{\unitlength}{1mm}
\begin{picture}(0,100)           
\put(10,-3){\begin{picture}(0,0) \includegraphics{ms9885.f4a} \end{picture}}
\put(65,-13){\begin{picture}(0,0) \includegraphics{ms9885.f4b} \end{picture}}
\put(90,58){\begin{picture}(0,0) \includegraphics{ms9885.f4c} \end{picture}}
\put(15,58){\begin{picture}(0,0) \includegraphics{ms9885.f4d} \end{picture}}
\end{picture}
\caption{The surface temperature map of FK~Com obtained for June 1997. Based on data from NOT. All the other parameters are as in Fig~\ref{map96r}.}
\label{map97}
\end{figure*}

\subsection{June-July 1996}
\label{r96}

The surface image and the fits to the observations of June-July 1996 are shown in Fig.~\ref{map96r}. The mean deviation of the spectroscopic observations from the model is 0.509~\%, which corresponds to an average S/N ratio of 196. The temperature range in the map is from 4294~K to 5792~K. The main feature in the map is an extended cool area centered on phases $0.4 - 0.5$ in the latitude interval $45^{\circ} - 62^{\circ}$. This feature has a temperature of $600 - 800$~K below the unspotted surface.

Photometric observations and the photometry calculated from the map are also shown in Fig.~\ref{map96r}. As can be seen, the B-V curve calculated from the map fits the simultaneous photometric observations quite well. For the V-curve the correlation is not as good. Nevertheless, the position of the light curve minimum at phase 0.5 is approximately reproduced. The poor correlation of the observed and calculated V-curves can be explained by the poor phase coverage and the small number of spectroscopic phases in the observations. Our tests show that removing the small cool spots at the phases 0.0--0.3 from the image makes the calculated light curve follow the observations better. These small spots are most likely artifacts caused by the noise in the observations and by the fact that only one spectrum has been taken during these phases. The noisy data and the poor phase coverage also explain the noisiness of the map and the higher spot temperatures than in the other maps. Due to the reasonable reproduction of the shape and position of the light curve minimum, as shown in our test, the surface position of the main feature in the map is considered reliable. The details of the shape of the spot, however, might not be exactly correct.

\subsection{July-August 1996}

The surface image and the fits to the observations of July-August 1996 are shown in Fig.~\ref{map96s}. The mean deviation of the spectroscopic observations from the model is 0.440~\%, which corresponds to an average S/N ratio of 227. The temperature range in the map is from 4021~K to 5605~K and the coolest feature has a temperature of $\sim1000$~K below that of the unspotted surface. The feature consists of two separate spots at phases 0.4 and 0.5 within the latitude interval $52^{\circ} - 62^{\circ}$ and a tail extending to lower latitudes.

The photometric observations used in the comparison are the same as in the case of the June-July 1996 map. Photometric observations and the photometry calculated from the map are also shown in Fig.~\ref{map96s}. As can be seen, the V and B-V curves calculated from the map fit the nearly simultaneous photometric observations very well. The light curve minimum is at phase 0.5. Light curves calculated from the map reproduce this minimum and the shapes of the observed light and colour curves. This implies that the average temperature of each phase is basically correct.

\subsection{April 1997}

The surface image and the fits to the observations of April 1997 are shown in Fig.~\ref{map97a}. The mean deviation of the spectroscopic observations from the model is 0.436~\%, which corresponds to an average S/N ratio of 229. The temperature range in the map is from 3689~K to 5518~K. Three cool features can be seen. The strongest ones are at phases 0.0 and 0.6 in the latitude interval $56^{\circ} - 73^{\circ}$. These features are $\sim1100$~K cooler than the unspotted surface. The third cool area is much weaker than the other two. It is situated at the phase 0.75 in the latitude interval $56^{\circ} - 68^{\circ}$ and it is $\sim500$~K cooler than the unspotted surface. 

V and B-V curves calculated from the map and the observations are also shown in Fig.~\ref{map97a}. The light curve minimum for this period is very broad and it spans almost 0.5 in phase, the minimum being approximately at the phase 0.75. Again, the photometry calculated from the map is supported by the observations. Some small differences in the position and the shape of the maximum can be seen in both V and B-V. The amplitude of the V curve has decreased significantly since July 1996.

\subsection{June 1997}

The surface image and the fits to the observations of June 1997 are shown in Fig.~\ref{map97}. The mean deviation of the spectroscopic observations from the model is 0.463~\%, which corresponds to an average S/N ratio of 216. The temperature range in the map is from 3709~K to 5640~K. The coolest feature in the map consist of two spots: the cooler one near phase 0.75 in the latitude interval of $40^{\circ} - 55^{\circ}$ has a temperature $\sim1100$~K below the unspotted surface and the weaker spot at phase 0.9 in the latitude interval $62^{\circ} - 73^{\circ}$ is $\sim600$~K cooler than the unspotted surface.

Photometry calculated from the map is compared with the simultaneous V and B-V observations. The light curves are also shown in Fig.~\ref{map97}. Also in this case the observed photometry is well reproduced by the calculations from the map. The light curve minimum is at phase 0.75. The amplitude of the V curve has decreased further.

\section{Discussion}

It seems that in the case of FK~Com the light curve minimum is always caused by several spots located in an extended active region. Our new surface maps allow us now to study the spot evolution and possible differential rotation in more detail.

\begin{figure*}
\setlength{\unitlength}{1mm}          
\begin{picture}(180,220) 
\put(17,0){\begin{picture}(0,0) \includegraphics{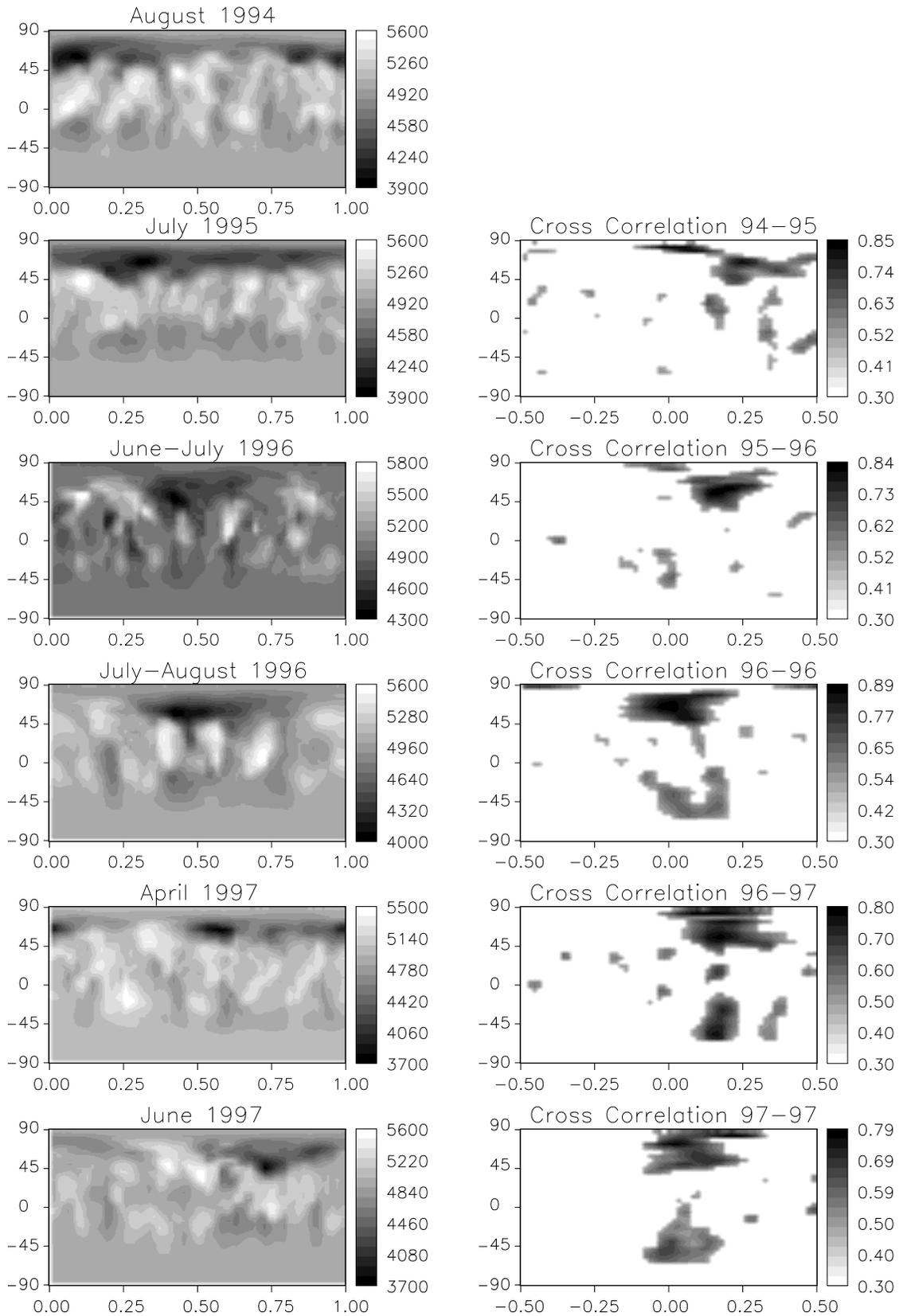} \end{picture}}
\end{picture}
\caption{Temperature maps for the years 1994 - 1997 with the results from the cross-correlation between consecutive images. In the plots the abscissa is phase and the ordinate is latitude in degrees. The maps for 1994 and 1995 are taken from Paper I.}
\label{cc}
\end{figure*}

\subsection{Spot evolution in 1994-1997}
\label{evo}

Fig.~\ref{cc} shows the temperature maps for the years 1994-1997 and the results from a cross-correlation between consecutive images. All the temperature maps show an extended active region, which can be resolved into two strong spots. The latitude range of the spots is quite stable from year to year, $45^{\circ}-70^{\circ}$. On the other hand, the longitude of the active region seems to move about 0.2 in phase within a year. 

The cross-correlation maps have been obtained by calculating the cross-correlation function for each individual latitude strip and then plotting them together as a map. As can be seen in the Fig.~\ref{cc}, the maximum of the cross-correlation function is obtained at the latitudes where the main spot features are located. In the cross-correlation procedure, each strip is individually normalized to the average fluctuations of that strip to get the proper value of the cross-correlation function. Due to this normalization in some images very small scale temperature variations are correlated and result in quite high values of cross-correlation function. This phenomenon is seen in some images as strong features at very low latitudes (between latitudes $-60^{\circ}$ and $-30^{\circ}$).

The cross-correlation confirms the migration rate to be on average $0.22\pm0.03$ in phase during one year. When calculating the average shift, the average of the cross-correlation function over all the latitudes with spots ($45^{\circ}$--$77^{\circ}$) was used. A 2nd order polynomial fit is then applied to determine the phase of the maximum of the cross-correlation function. The cross-correlation between April 1997 and June 1997 was excluded from the measurements, because it lead to a much larger error in the result, $0.24\pm0.09$. This is most likely due to a short-term rapid evolution of the active region during spring 1997. Our photometry from the end of 1997 shows a very different light-curve shape and behaviour than in early 1997 (Korhonen \cite{kor2}). This might be caused by the so-called 'flip-flop' phenomenon, where the longitude of the active region switches by $180^{\circ}$ (cf.\ Jetsu \cite{jetsu4}). The rapid evolution during spring 1997 might indicate the start of a new 'flip-flop'.

The spot evolution on the shorter time scales can be studied with the four maps obtained in 1996 and 1997. In the cross-correlation between the June-July 1996 and July-August 1996 maps it can be seen that these two maps are similar and no significant differences between them is revealed. The only difference is the small phase shift of the active region. This phase shift is about 0.016 during the 30 day period. The shift has been determined from all the latitudes between $45^{\circ}$ and $77^{\circ}$ using a 2nd order polynomial fit to the values of the cross-correlation function. The shift obtained is in good agreement with the migration predicted from the shift of the active region during the four year period (0.019 in phase in 30 days). Similarly, the cross-correlation between April and June 1997 maps shows a phase shift of 0.063, which is to be compared with the predicted shift of 0.045 within 72 days.

The detected migration of the active region is probably caused by a difference between the adopted rotation period and the real period. If this is true then the real period is $2\fd4037\pm 0.0005$ (using additionally the cross-correlation between April 1997 and June 1997 gives $2\fd4040\pm 0.0014$). Due to the solid body rotation of the stellar surface within the spotted latitudes (see Sect.~\ref{diff}), this value can be taken to be representative of the stellar rotation. The stability of the active region during a four year period and the very similar substructures within it indicate that the inversion method used in this work is quite reliable and that the active regions in FK~Com are stable between the anticipated flip-flop's.

\subsection{Differential rotation}
\label{diff}

\begin{figure}
\setlength{\unitlength}{1mm}          
\begin{picture}(0,92) 
\put(5,0){\begin{picture}(0,0) \includegraphics{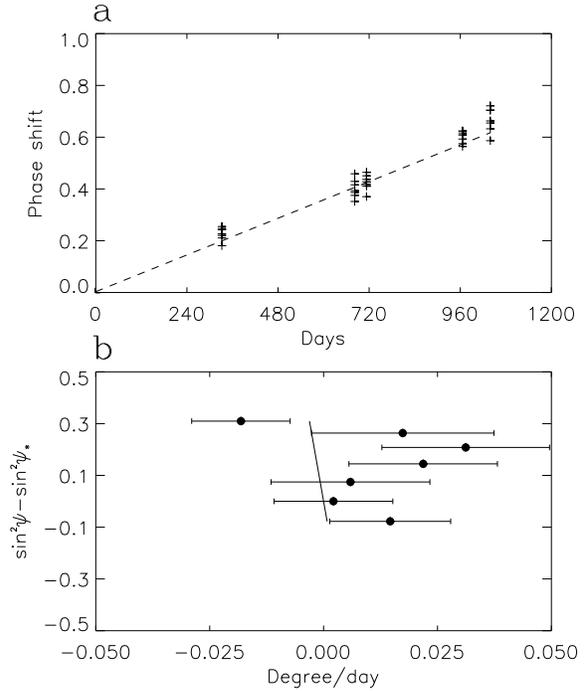} \end{picture}}
\end{picture}
\caption{{\bf a} Phase shifts relative to the August 1994 map measured from all six maps obtained between 1994 and 1997 and a fit to the measurements at the comparison latitude $\psi_{\ast}=51.75^{\circ}$. {\bf b} The average phase shifts in degrees per day plotted against $\sin^{2}(\psi)-\sin^{2}(\psi_{\ast})$, after the first degree polynomial fit to $\psi_{\ast}=51.75^{\circ}$ has been removed. The error bars indicate the standard deviation. Also the weighted fit of the function $f(\psi)=\beta(\sin^{2}(\psi)-\sin^{2}(\psi_{\ast}))$ to the measurements is shown as a straight line, yielding $\beta=-0.01\pm0.03$.}
\label{difrot}
\end{figure}

The results from the cross-correlation were also used for checking the possibility of surface differential rotation. Measurements from all the maps and all the latitudes with spots ($45^\circ-77^{\circ}$) were used to determine the possible differential rotation described by the parameter $\alpha$ as follows: 
\begin{equation}
\alpha=\frac{\Omega_{0}-\Omega_{\rm p}}{\Omega_{0}},
\end{equation}
where $\Omega_{0}$ and $\Omega_{\rm p}$ are the equatorial and polar angular velocities, respectively. Unfortunately, there are no spots near the equator. So, a comparison latitude ($\psi_{\ast}$) other than the equator must be used. With a $\sin^{2}\psi$ law the angular velocity at any latitude can be expressed as:
\begin{equation}
\Omega=\Omega_{0}+\beta\sin^{2}\psi,
\label{eq:omega}
\end{equation}
where $\psi$ is the latitude and $\Omega$ the angular velocity at this latitude. The polar angular velocity can be obtained from Eq.~\ref{eq:omega} and $\alpha$ can then be represented as:
\begin{equation}
\alpha=\frac{-\beta}{\Omega_{0}}.
\end{equation}

The value of $\beta$ can be determined from the measurements of the longitudinal shifts at different latitudes. These are expressed in phases and are plotted in the upper panel of Fig.~\ref{difrot}. The shifts comprise the annual migration plus a possible differential rotation. To remove the annual migration, we have chosen the lowest latitude which always has spots in our maps as the comparison latitude ($\psi_{\ast}=51.75^{\circ}$) and subtracted the average shift at this latitude (dashed line in the plot) from the measurements.

Then, the residual shift in degrees per day for each latitude can be expressed as:
\begin{equation}
f(\psi)=\Omega(\psi)-\Omega(\psi_{\ast})=\beta(\sin^{2}(\psi)-\sin^{2}(\psi_{\ast})).
\end{equation} 
They are plotted in the lower panel of Fig.~\ref{difrot}. In the plot the standard deviations are shown as error bars. Then $\beta$ can be determined by using a first order polynomial fit with the inverse of the variance of the average in each point as the weight. The fit resulted in $\beta=-0.01\pm0.03$. For the calculation of $\alpha$ the angular velocity at the equator is still needed. That can be calculated from Eq.~(\ref{eq:omega}), after the velocity at the comparison latitude is known. For $\psi_{\ast}=51.75^{\circ}$ we get $P=2\fd4037\pm 0.0009$, which gives $\Omega_{\ast}=149.769^{\circ}/{\rm day}$. Together with $\beta$ this yields $\alpha=0.0001\pm0.0002$. This result is consistent with solid body rotation of the stellar surface. For further pinpointing differential rotation, more maps extending the time span would be needed.

\subsection{Chromospheric activity and the spots}

\begin{figure}
\setlength{\unitlength}{1mm}          
\begin{picture}(0,100) 
\put(5,0){\begin{picture}(0,0) \includegraphics{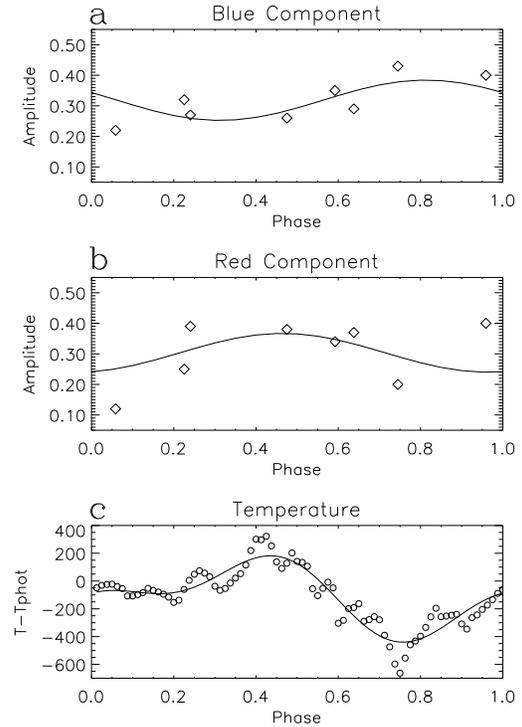} \end{picture}}
\end{picture}
\caption{Emission in the blue ({\bf a}) and red wing ({\bf b}) of the H$\alpha$-line (observations from Oliveira \& Foing \cite{oliv}) and the average photospheric temperature ({\bf c}) determined from the June 1997 map using the latitudes between $36^{\circ}$ and $72^{\circ}$. The amplitude for H$\alpha$ emission is given in continuum units. In the two upper plots the lines are the first order harmonic fits and in the lower plot a second order harmonic fit.}
\label{chromo}
\end{figure}

Oliveira \& Foing (\cite{oliv}) analysed the chromospheric activity of FK~Com for May-June 1997, covering partly the same period as our June 1997 temperature map. At two phases a flare had been identified by Oliveira \& Foing (\cite{oliv}). H$\alpha$ observations at these phases were not used when comparing with the June 1997 map.

To search for a correlation of the chromospheric activity with the photospheric spots, we use the measurements of residual emission in the blue and red wings of H$\alpha$-line (at radial velocities of $\pm200$\kms). The measurements are plotted in Figs.~\ref{chromo}a and \ref{chromo}b together with a 1st degree harmonic fit. Fig.~\ref{chromo}c shows the average temperature distribution in the June 1997 map calculated from the spotted latitudes, $36^{\circ}$--$72^{\circ}$, and a 2nd degree harmonic fit. From the comparison we see that the emission in the blue wing is the strongest at the phase of the photospheric active region, while the maximum of the emission in the red wing is shifted by approximately 0.5 in phase from that. We also note that the maximum and the amplitude of the H$\alpha$ emission has the same value in both wings. These observations can be interpreted with a simple model of constant emission from the chromosphere, seen during the minimum of the emission, and matter flowing up from the active region and resulting in variable rotationally modulated emission. When the spots are at the centre of the disk the flow is strongest towards the Earth, so the largest emission is seen in the blue wing. Half a rotation later the flow is pointed away and the strongest emission is seen in the red wing. This model suggests hot active plages in the chromosphere in the vicinity of the spots with outflows of about 200\kms. The similar amount of emission in the blue and red wings during the minima implies a more-or-less constant emission from the chromosphere.

The limited number of H$\alpha$ observations makes these results very preliminary. Having obtained observations simultaneously in H$\alpha$ and the spectral region used for surface imaging for the years 1998--1999, we intend to study this possible correlation in more detail later.

\section{Conclusions}

We have obtained four new temperature maps of FK~Com for the years 1996 and 1997. These new maps are compared with the previously published maps for 1994 and 1995. The following conclusions can be drawn:

\begin{enumerate}
\item The active region is stable for at least 4 years.
\item The active region moves on average $0.22\pm 0.03$ in phase within a year. This migration is probably caused by a difference between the adopted rotation period and the real period; if this is true then the real rotation period is $\sim 2\fd4037\pm 0.0005$.
\item No evidence for surface differential rotation could be found from temperature maps separated by 30 days, 72 days and about a year. Assuming a $\sin^{2}\psi$ law, measurements from latitudes between $45^\circ$ and $77^{\circ}$ give $\alpha=0.0001\pm0.0002$, where $\alpha$ is the difference between polar and equatorial angular velocities relative to the equatorial angular velocity. This result indicates a solid body rotation of the stellar surface.
\item Comparison between our June 1997 temperature map and the chromospheric activity observed in H$\alpha$ suggests the presence of hot plages in the chromosphere in the vicinity of the spots with outflows of about 200\kms. More observations of chromospheric lines, simultaneous with the photospheric lines used for surface imaging, would be needed to confirm the model.
\end{enumerate}

\begin{acknowledgements}
We thank I.\ Ilyin and R.\ Duemmler for their help in obtaining the observations from NOT and Rozhen. We thank R.\ Duemmler also for his careful reading of this manuscript. We would also like to thank our referee J.B. Rice for his comments. This research was partly supported by the EC Human Capital and Mobility (Network) grant ``Late type stars: activity, magnetism, turbulence'' No.\ ER-BCHRXCT940483. This project has been supported by the European Commission through the Activity "Access to Large-Scale Facilities" within the Programme "Training and Mobility of Researchers" awarded to the Instituto de Astrof{\'i}sica de Canarias to fund European Astronomers' access to its Roque de Los Muchachos and Teide Observatories (European Northern Observatory), in the Canary Islands. The work of HK was supported by the Finnish graduate school in Astronomy and Space Physics and by grants from the Jenny and Antti Wihuri foundation, the Vilho, Yrj{\"o} and Kalle V{\"a}is{\"a}l{\"a} foundation and the Nordic Optical Telescope Scientific Association. TH was supported by the Jenny and Antti Wihuri foundation and by Helsinki University research funds for the project ``Time series analysis in astronomy'' (No.\ 974/62/98). KGS acknowledges the receipt of FWF grants S7301 and S7302 from the Austrian Science Foundation. The calculations for the inversions were carried out on the Cray C94/128 supercomputer at the Centre for Scientific Computing (Espoo, Finland). Nordic Optical Telescope is operated on the island of La Palma jointly by Denmark, Finland, Iceland, Norway, and Sweden, in the Spanish Observatorio del Roque de los Muchachos of the Instituto de Astrofisica de Canarias.
\end{acknowledgements}

\end{document}